\documentclass[11pt,twocolumn]{article}

\usepackage{fix-cm}
\usepackage[T1]{fontenc}
\usepackage[margin=1in]{geometry}
\usepackage{amsmath,amssymb}
\usepackage{booktabs}
\usepackage{tabularx}
\usepackage{listings}
\usepackage{xcolor}
\usepackage{hyperref}
\usepackage{graphicx}
\usepackage{caption}
\usepackage{subcaption}
\usepackage{enumitem}
\usepackage{titlesec}
\usepackage{fancyhdr}
\usepackage{parskip}
\usepackage{tikz}
\usetikzlibrary{shapes.geometric, arrows.meta, positioning, fit, backgrounds}
\usepackage{array}
\usepackage{multirow}
\usepackage{url}
\usepackage[numbers]{natbib}
\usepackage{pgfplots}
\pgfplotsset{compat=1.18}

\definecolor{codeblue}{rgb}{0.08,0.38,0.74}
\definecolor{codegray}{rgb}{0.5,0.5,0.5}
\definecolor{codegreen}{rgb}{0.13,0.55,0.13}
\definecolor{codebg}{rgb}{0.97,0.97,0.97}
\definecolor{accentblue}{rgb}{0.08,0.38,0.74}

\lstdefinestyle{typescript}{
  backgroundcolor=\color{codebg},
  commentstyle=\color{codegreen}\itshape,
  keywordstyle=\color{codeblue}\bfseries,
  stringstyle=\color{orange},
  basicstyle=\ttfamily\footnotesize,
  breakatwhitespace=false,
  breaklines=true,
  keepspaces=true,
  numbers=left,
  numbersep=5pt,
  numberstyle=\tiny\color{codegray},
  showspaces=false,
  showstringspaces=false,
  tabsize=2,
  frame=single,
  framerule=0.4pt,
  rulecolor=\color{codegray},
}
\titleformat{\section}{\large\bfseries\color{accentblue}}{\thesection.}{0.5em}{}
\titleformat{\subsection}{\normalsize\bfseries}{\thesubsection.}{0.5em}{}
\titleformat{\subsubsection}{\normalsize\itshape}{\thesubsubsection.}{0.5em}{}

\hypersetup{
  colorlinks=true,
  linkcolor=accentblue,
  citecolor=accentblue,
  urlcolor=accentblue,
  pdftitle={Internal APIs Are All You Need},
  pdfauthor={Lewis Tham, Nicholas Mac Gregor Garcia, Jungpil Hahn},
}

\begin{document}

\title{\textbf{Internal APIs Are All You Need}\\[6pt]
\large\textit{Shadow APIs, Shared Discovery, and the Case Against Browser-First Agent Architectures}}

\author{%
  Lewis Tham\textsuperscript{1} \quad
  Nicholas Mac Gregor Garcia\textsuperscript{2} \quad
  Jungpil Hahn\textsuperscript{2} \\[6pt]
  \textsuperscript{1}Unbrowse AI \quad \texttt{lewis@unbrowse.ai} \\
  \textsuperscript{2}School of Computing, National University of Singapore \\
  \texttt{ngarcia@nus.edu.sg} \quad \texttt{jungpil@nus.edu.sg}
}

\date{}
\maketitle
\thispagestyle{fancy}

\begin{abstract}
Autonomous agents increasingly interact with the web, yet most websites remain designed for human browsers --- a fundamental mismatch that the emerging ``Agentic Web'' must resolve. Agents must repeatedly browse pages, inspect DOMs, and reverse-engineer callable routes --- a process that is slow, brittle, and redundantly repeated across agents. We observe that every modern website already exposes internal APIs (sometimes called \emph{shadow APIs}) behind its user interface --- first-party endpoints that power the site's own functionality. We present Unbrowse, a shared route graph that transforms browser-based route discovery into a collectively maintained index of these callable first-party interfaces. The system passively learns routes from real browsing traffic and serves cached routes via direct API calls. In a single-host live-web benchmark of equivalent information-retrieval tasks across 94 domains, fully warmed cached execution averaged 950\,ms versus 3{,}404\,ms for Playwright browser automation (3.6$\times$ mean speedup, 5.4$\times$ median), with well-cached routes completing in under 100\,ms. A three-path execution model --- local cache, shared graph, or browser fallback --- ensures the system is voluntary and self-correcting. A three-tier micropayment model via the x402 protocol charges per-query search fees for graph lookups (Tier~3), a one-time install fee for discovery documentation (Tier~1), and optional per-execution fees for site owners who opt in (Tier~2). All tiers are grounded in a necessary condition for rational adoption: an agent uses the shared graph only when the total fee is lower than the expected cost of browser rediscovery.

\end{abstract}

\section{Introduction}

The web was built for humans to navigate pages. Agents do not want pages --- they want actions. Today, a web-capable agent repeatedly pays a \emph{discovery tax} to interact with any website: it must open the site, inspect the DOM, click through the UI, infer requests, retry when the workflow breaks, and spend LLM tokens reasoning about page state. This process is not only expensive but \emph{redundantly} so, as different agents re-derive the same workflows on the same sites, each paying the full discovery cost independently.

Current approaches to web interaction fall into two categories, each with significant limitations:

\begin{enumerate}[leftmargin=*]
  \item \textbf{Browser automation:} Systems like WebVoyager~\cite{he2024webvoyager}, AutoWebGLM~\cite{lai2024autowebglm}, and BrowserAgent~\cite{chen2025browseragent} use LLMs to navigate and interact with web interfaces through browser actions. While effective for human-designed workflows, these approaches are inherently slower than direct API access, brittle (breaking when the UI changes), and computationally expensive; Song et al.~\cite{song2024beyond} show that agents with API access outperform browsing-only agents on realistic web benchmarks.

  \item \textbf{Official APIs:} Where available, APIs provide clean machine interfaces. However, coverage is limited --- most websites lack public APIs --- and access often requires registration, approval, and compliance with rate limits.
\end{enumerate}

We propose a third path: Unbrowse, a \emph{shared route graph} that turns repeated browser discovery into shared memory. Rather than treating every interaction as a fresh reverse-engineering exercise, Unbrowse passively learns callable interfaces from real usage and stores them in a shared index. Agents remain free to rediscover routes themselves --- the shared graph competes directly against self-production and wins only when it delivers genuine surplus (Section~\ref{sec:eval}).
In this sense, Unbrowse implements a concrete layer of what Yang et al.\ term the \emph{Agentic Web}~\cite{yang2025agenticweb} --- an infrastructure where autonomous agents interact with digital services directly rather than through human-designed interfaces.

This paper makes the following contributions:

\begin{enumerate}[leftmargin=*]
  \item A \textbf{shared route graph architecture} that passively learns callable web interfaces from real browsing traffic, transforming redundant private discovery into a collectively maintained commons.

  \item A \textbf{conceptual economic model} grounded in a necessary adoption condition --- $f_{\text{route}} < c_{\text{rediscovery}}$ --- creating a market-disciplined system with a built-in outside option.

  \item A \textbf{novel delta-based attribution mechanism} for route-level micropayments via the x402 protocol~\cite{coinbase2025x402}, compensating contributors in proportion to their marginal contribution to route quality.

  \item \textbf{Empirical evaluation} across 94 domains demonstrating significant speedups over browser automation on equivalent information-retrieval tasks, with analysis of both warmed-cache and cold-start performance.
\end{enumerate}

We situate Unbrowse within four areas of prior work: web agents and browser automation, tool learning and API discovery, API aggregation platforms, and agent economies and communication protocols.

\section{Related Work}

\subsection{Web Agents and Browser Automation}

The field of web agents has evolved from early text-based systems to modern multimodal approaches. WebGPT~\cite{nakano2021webgpt} demonstrated browser-assisted question answering, while ReAct~\cite{yao2023react} introduced a more general reasoning-and-action framework that influenced many later interactive agent designs. Contemporary systems such as WebVoyager~\cite{he2024webvoyager}, together with benchmarks such as Mind2Web~\cite{deng2023mind2web}, show rapid progress in web agents while also highlighting the fundamental efficiency constraints of GUI-based interaction.

Song et al.~\cite{song2024beyond} directly addressed this limitation in ``Beyond Browsing: API-Based Web Agents,'' demonstrating that hybrid agents with API access outperform pure browsing agents by over 24\% on WebArena benchmarks~\cite{zhou2024webarena}. This provides empirical validation for our core thesis: direct API-style access is superior to browser automation when available. Our contribution extends this insight by asking what happens when such access can be collectively discovered and shared, rather than requiring pre-documented APIs.

\subsection{Tool Learning and API Discovery}

The integration of external tools into LLM workflows has emerged as a critical capability. Toolformer~\cite{schick2023toolformer} pioneered self-supervised tool learning, while subsequent work has explored hierarchical tool retrieval~\cite{du2024anytool}, natural-language tool calling interfaces~\cite{johnson2025nltool}, and automatic tool generation~\cite{shi2025toolwild}. Gorilla~\cite{patil2023gorilla} demonstrated that LLMs fine-tuned on API documentation can generate accurate API calls, and ToolBench~\cite{qin2023toolbench} introduced a large-scale benchmark for tool-augmented LLMs spanning over 16{,}000 real-world APIs. Prior work in network traffic analysis --- tools such as mitmproxy~\cite{mitmproxy2024} and HAR-to-OpenAPI converters built on the W3C HAR specification~\cite{w3c2012har} --- has long enabled developers to inspect and document internal APIs.

Our approach extends both lines of work. The novelty lies not in traffic capture per se, but in the synthesis of passive private observation into a shared index: where prior tools produce developer artefacts for human use, Unbrowse produces a collectively maintained index of callable interfaces for autonomous agent consumption.

\subsection{API Aggregation Platforms}

Commercial API marketplaces such as RapidAPI and Postman's public workspace have demonstrated the value of centralised API discovery. RapidAPI aggregates over 40{,}000 official APIs behind a unified gateway with standardised authentication and billing. Postman's public API network provides searchable collections of documented endpoints. However, these platforms are limited to \emph{official}, publicly documented APIs --- they cannot address the long tail of websites that expose no public API. Our approach differs in two respects: routes are discovered automatically from browser traffic rather than manually submitted by API providers, and the graph covers internal APIs that were never intended for public consumption but are nonetheless callable.

In parallel to shared discovery approaches, recent work has explored making websites explicitly agent-ready. webMCP~\cite{webmcp2025} proposes AI-native client-side interaction patterns that expose machine-usable interfaces directly from the website itself. This approach is complementary to Unbrowse: webMCP describes how future sites may be designed for agents \emph{ex ante}, whereas Unbrowse targets the existing web \emph{ex post} by passively discovering and caching callable internal interfaces on sites that were never built for agent-native access. Table~\ref{tab:comparison} provides a qualitative comparison.

\begin{table}[h!]
\centering
\caption{Qualitative comparison with existing API aggregation platforms.}
\label{tab:comparison}
\scriptsize
\setlength{\tabcolsep}{3pt}
\begin{tabular}{@{}lccc@{}}
\toprule
& \textbf{Rapid-} & \textbf{Post-} & \textbf{Un-} \\
\textbf{Property} & \textbf{API} & \textbf{man} & \textbf{browse} \\
\midrule
Coverage    & Official & Official & Any site \\
Discovery   & Manual   & Manual   & Passive \\
Internal APIs & \texttimes & \texttimes & \checkmark \\
Agent-native  & Partial  & \texttimes & \checkmark \\
\bottomrule
\end{tabular}
\end{table}

\subsection{Agent Economies and Communication Protocols}

Recent work has formalised interoperability and communication protocols --- including MCP (Model Context Protocol), A2A (Agent-to-Agent), and ANP (Agent Network Protocol) --- for connecting agents to tools and to one another~\cite{ehtesham2025survey,singh2025evolution,tran2025multiagent}. These developments build on a longer history of multi-agent systems research~\cite{ferber1999multiagent} and broader narratives linking the Semantic Web and MAS traditions to today's Agentic Web~\cite{petrova2025semantic}. These protocols define \emph{how} agents connect to tools but do not address \emph{what} is available to connect to. Unbrowse complements these protocols by populating the registry of callable interfaces that agents discover through them.

A shared route graph only works if agents continuously contribute to it --- discovering new routes for new sites and verifying the accuracy of existing routes on an ongoing basis. Without sustained contribution, the index degrades as APIs drift and coverage stagnates. This is fundamentally an incentive design problem: how to make discovery and maintenance economically attractive.

Yang et al.~\cite{yang2025agenticweb} articulate the vision of an ``Agentic Web'' where autonomous systems interact directly with digital infrastructure. Vaziry et al.~\cite{vaziry2025multiagent} integrate ledger-anchored identities with the x402 micropayment protocol~\cite{coinbase2025x402} to enable trustless economic interactions between agents. ERC-8004~\cite{erc8004} proposes on-chain registries for agent identity and reputation. The shared route graph can be understood as a knowledge commons in the sense of Ostrom~\cite{ostrom1990governing}: a collectively maintained resource whose value depends on sustained contribution and ongoing maintenance. The economic model described in Section~\ref{sec:econ} draws on this framing --- the three-path execution model ensures the commons competes with private production (browser rediscovery), providing what Ostrom calls the ``exit option'' that disciplines the system and prevents free-riding.

Collectively, prior work has articulated the vision, protocols, and governance frameworks for an Agentic Web~\cite{yang2025agenticweb,lu2025build,betaweb2025}, but empirical implementations of shared discovery for callable web interfaces remain limited. Unbrowse contributes one such implementation: a deployed system that targets the existing web by letting agents interact with web services through collectively discovered interfaces rather than only through human-designed pages, and evaluating that approach across 94 live domains.

\section{System Design}

\subsection{System Architecture}

Unbrowse serves as a drop-in replacement for browser-based agent interaction. It integrates with any agent host that supports CLI tools, MCP (Model Context Protocol), or the AgentSkills.io skill standard~\cite{agentskills2025}.\footnote{Unbrowse has been tested with Claude Code, OpenClaw, Codex, Cursor, and Windsurf at time of writing.}

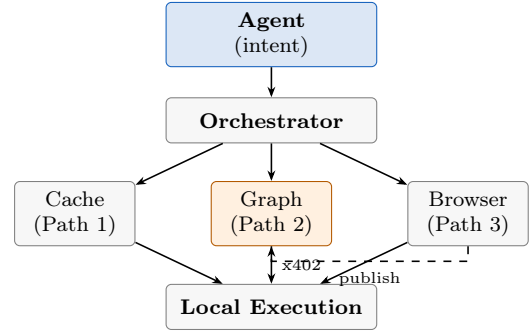
\begin{figure}[h!]
\centering
\begin{tikzpicture}[
  node distance=0.35cm,
  every node/.style={font=\scriptsize},
  box/.style={draw, rounded corners=2pt, minimum width=2cm,
              minimum height=0.6cm, align=center, font=\scriptsize},
  agentbox/.style={box, fill=accentblue!15, draw=accentblue},
  sysbox/.style={box, fill=codebg, draw=codegray},
  econbox/.style={box, fill=orange!12, draw=orange!70!black},
  arr/.style={-{Stealth[length=4pt]}, semithick},
  darr/.style={-{Stealth[length=4pt]}, semithick, dashed},
]
  \node[agentbox, minimum width=2.8cm] (agent) {\textbf{Agent}\\(intent)};
  \node[sysbox, below=0.4cm of agent, minimum width=2.8cm] (orch) {\textbf{Orchestrator}};

  \node[sysbox, below left=0.5cm and 0.4cm of orch, minimum width=1.6cm] (cache) {Cache\\(Path 1)};
  \node[econbox, below=0.5cm of orch, minimum width=1.6cm] (market) {Graph\\(Path 2)};
  \node[sysbox, below right=0.5cm and 0.4cm of orch, minimum width=1.6cm] (browser) {Browser\\(Path 3)};

  \node[sysbox, below=0.5cm of market, minimum width=2.8cm] (exec) {\textbf{Local Execution}};

  \draw[arr] (agent) -- (orch);
  \draw[arr] (orch) -- (cache);
  \draw[arr] (orch) -- (market);
  \draw[arr] (orch) -- (browser);
  \draw[arr] (cache) -- (exec);
  \draw[arr] (market) -- node[right, font=\tiny] {x402} (exec);
  \draw[arr] (browser) -- (exec);
  \draw[darr] (browser.south) -- ++(0,-0.2) -| node[below, font=\tiny, pos=0.25] {publish} (market.south);
\end{tikzpicture}
\caption{Three-path execution model. The orchestrator routes requests to the local cache, shared graph (x402 payment), or browser fallback. Discovered routes are published back (dashed).}
\label{fig:arch}
\end{figure}

The architecture consists of two integrated layers: a \emph{Capability Layer} responsible for passive route discovery and shared graph maintenance (described in this section), and an \emph{Economic Layer} handling route-level pricing and contributor payouts (detailed in Sections~\ref{sec:econ} and~5). Figure~\ref{fig:arch} illustrates the three-path execution model and the interaction between these layers.

When an agent asks Unbrowse to perform an action on a website, the orchestrator routes the request automatically through a priority chain. First (\textbf{Path~1}), it checks the local route cache for a recent hit. If no cached route is available, it queries the shared graph (\textbf{Path~2}) for a matching skill; if found, the skill is installed locally and executed via direct API calls --- no browser is launched, no DOM is parsed, and no LLM tokens are spent reasoning about page state. If a skill exists but fails, automatic fallback to browser capture and re-learning occurs transparently.

If no suitable route exists in either cache or graph (\textbf{Path~3}), Unbrowse falls back to Kuri~\cite{pradhan2025kuri}, a Zig-native CDP broker that replaces heavyweight browser automation frameworks. In our implementation measurements, Kuri's packaged broker is substantially smaller and faster to start than a full Playwright runtime, though these figures should be understood as implementation-specific rather than independently benchmarked claims. Kuri launches headless Chrome, captures network traffic, reverse-engineers the API endpoints behind the UI, and publishes the learned skill to the shared graph for future reuse.

From the agent's perspective, nothing changes: it describes what it wants in natural language, and Unbrowse handles routing transparently. Every time any agent discovers a new route, every other agent benefits. The graph is not mined through blind exploration but distilled from real demand --- coverage naturally concentrates on the sites and workflows that agents actually need, analogous to cooperative caching in CDN systems~\cite{wolman1999cooperative}. Crucially, agents can always bypass the shared graph and fall back to browser rediscovery; the graph wins only when it creates genuine value.

Unbrowse is distributed as an npm package that installs the CLI, bootstraps the Kuri browser runtime, and registers the tool with any detected agent host. It generates a \texttt{SKILL.md} file following the AgentSkills.io open standard~\cite{agentskills2025} that any skill-compatible agent host can consume directly. Agents interact with Unbrowse through the CLI or its local HTTP API (\texttt{POST /v1/intent/resolve}). All credentials remain in a local encrypted vault; published skills contain only endpoint documentation and schemas, never credentials.

\subsection{Capability Layer: Route Discovery Pipeline}

As agents browse and complete tasks through Unbrowse, the system passively observes network traffic and infers reusable route abstractions, building on established techniques in network traffic analysis~\cite{mitmproxy2024} and extending them into an agent-native pipeline. Discovered routes are published to a cloud-hosted marketplace backed by semantic vector search.

\subsubsection{Traffic Observation and Filtering}

Raw browser traffic contains significant noise from analytics, advertising networks, and CDN requests for static assets. We employ heuristic filtering based on Content-Type analysis (JSON/XML responses indicate API calls), URL pattern matching, request method analysis (POST/PUT/PATCH are strong API indicators), and response structure analysis.

\subsubsection{Route Extraction and Normalisation}

From filtered traffic, the system extracts endpoints and their parameter structures, response shapes, authentication assumptions, and route-specific metadata. These are converted into pointer-like route abstractions --- not raw web content, but a maintained map of callable interfaces.

\subsubsection{What the Graph Stores}

The shared graph is primarily an index of pointers to callable interfaces, not a storage layer for the raw web. It stores route definitions and schemas, parameter structures and response shapes, and authentication descriptors. Derived attributes --- confidence (computed from execution success rates), freshness (inverse decay from last verification date), and health (updated by the periodic verification loop described below in Skill Lifecycle and Schema Drift) --- are recomputed continuously from execution telemetry rather than stored statically.

\subsubsection{Skill Packaging}

Extracted routes are packaged into skills following the AgentSkills.io open standard~\cite{agentskills2025} --- structured capability bundles that any compatible agent host can consume. Each skill contains:
\begin{itemize}[leftmargin=*]
  \item A \texttt{SKILL.md} file providing human-readable endpoint documentation, compatible with any skill-aware agent host.
  \item An \texttt{auth.json} file holding captured authentication credentials that are encrypted and stored locally (never published to the marketplace).
  \item An \texttt{api.ts} file containing a generated TypeScript client for programmatic access.
\end{itemize}
Skills are the unit of sharing: what gets published to the marketplace is the knowledge of how to interact with a site, not credentials or execution capability. A single website may yield multiple skills (one per discovered workflow or endpoint group), and different agents may discover different endpoints on the same site. Skills are merged at the domain level when overlapping endpoints are detected, with attribution tracked per-endpoint.

\subsection{Composite Scoring and Intent Resolution}

When an agent requests a task, the orchestrator follows a priority chain. First, it checks the \emph{route cache} (24-hour TTL, persisted to disk) for an instant hit if the same intent was recently resolved. If no cache hit is found, the system performs a \emph{marketplace search} using semantic vector search ranked by a composite score comprising 40\% embedding similarity, 30\% reliability, 15\% freshness, and 15\% verification status. These weights were set heuristically to prioritise semantic relevance while giving substantial weight to demonstrated reliability; they are configurable parameters, and adapting them from execution telemetry is planned future work. If no marketplace result is suitable, \emph{live capture} launches a headless browser to record network traffic, reverse-engineer API endpoints, and publish a new skill. As a final fallback for static or server-side-rendered sites where no API endpoints are found, structured data is extracted from rendered HTML.

\subsection{Skill Lifecycle and Schema Drift}

Skills in the marketplace follow a natural lifecycle driven by real usage data. \emph{Active} skills are published, queryable, and executable. \emph{Deprecated} skills have triggered low-reliability warnings and remain discoverable but are ranked lower. \emph{Disabled} skills have confirmed endpoint failures and are removed from search results until re-verified. A background verification loop runs every 6 hours on each agent's local Unbrowse server, executing safe GET endpoints to detect failures and schema drift. The type system includes a \texttt{consecutive\_failures} counter for automatic deprecation; wiring this counter to a threshold-based deprecation policy is in progress.

Web APIs change without notice. The system continuously monitors for schema drift by comparing live response structures against documented response schemas. When critical drift is detected --- removed fields, type changes --- the affected endpoint is flagged for re-verification. This ensures the graph reflects current API reality rather than stale documentation.

\section{Economic Model}\label{sec:econ}

\subsection{The Adoption Condition}

The shared route graph replaces browser sessions with direct API calls for cached lookups. A cost-minimising agent will use the shared graph if and only if:
\begin{equation}
  f_{\text{route}} < c_{\text{rediscovery}}
  \label{eq:foundational}
\end{equation}
where $f_{\text{route}}$ denotes the total fee an agent incurs through the shared graph (decomposed into Tier~1, 2, and~3 components in Section~5), and the expected rediscovery cost is decomposed as:
\begin{equation}
  c_{\text{rediscovery}} = c_{\text{latency}} + c_{\text{compute}} + c_{\text{tokens}} + p_{\text{fail}} \cdot c_{\text{retry}}
  \label{eq:rediscovery}
\end{equation}

Here $c_{\text{latency}}$ is the opportunity cost of multi-second page loads and sequential interaction overhead in browser-based agents, $c_{\text{compute}}$ is browser runtime cost (approximately 500\,MB RAM per instance), $c_{\text{tokens}}$ is LLM reasoning tokens for DOM interpretation and navigation, $p_{\text{fail}}$ is browser-task failure probability on realistic web benchmarks~\cite{zhou2024webarena,song2024beyond}, and $c_{\text{retry}}$ is the cost of retrying a failed interaction.

This is a \emph{necessary condition} for rational adoption, not a complete market model. It establishes the ceiling: no rational agent will pay $f_{\text{route}} \geq c_{\text{rediscovery}}$, so the shared graph is disciplined by a real outside option. Section~\ref{sec:eval} provides empirical estimates for both sides of the inequality.

\subsection{What Agents Pay For}

Agents do not pay for access to the web itself. They pay for \emph{speed}. Sub-second execution transforms what agents can do: multi-site research workflows that take minutes through browser automation complete in seconds through the marketplace. This speed advantage is categorical, not incremental --- it determines whether agents can chain web interactions into real-time workflows or whether every web task becomes a blocking bottleneck.

Secondarily, agents also pay for lower compute burn (no browser runtime), lower LLM token burn (no reasoning about page state), lower failure risk, and maintained route knowledge kept fresh by the network.

The adoption condition as stated omits the marketplace search cost. In practice, each graph query incurs a Tier~3 search fee (Section~5.3) covering the semantic vector lookup (typically 50--200\,ms latency, \$0.001--0.005 per query). For a single interaction the Tier~3 fee is negligible relative to the multi-second browser alternative; however, for agents that query the graph frequently, cumulative Tier~3 spend must be included in the adoption calculus. A more complete model would express the condition as $f_{\text{search}} + f_{\text{install}} + n \cdot f_{\text{exec}} < c_{\text{rediscovery}}$ where $n$ is the number of executions on Tier~2 sites.

\subsection{Why a Shared Commons Can Emerge}

Following Ostrom's framework for knowledge commons~\cite{ostrom1990governing}, a shared route registry makes economic sense when many actors repeatedly rediscover similar routes, when pooled usage improves route quality faster than isolated usage, when maintenance is ongoing and expensive, and when duplication of discovery work is wasteful. The three-path execution model provides what Ostrom calls the ``exit option'' --- competitors can always rebuild privately, and this credible threat disciplines the commons. The shared index wins only when it is cheaper than repeated rediscovery and when pooled usage improves coverage faster than isolated efforts.

We emphasise that this is a \emph{conceptual} economic framework. Formal equilibrium analysis --- including conditions under which the commons is underprovided, pricing strategies beyond the ceiling constraint, and welfare characterisation --- remains future work. The current contribution is the architectural design that makes such a commons technically feasible, together with the empirical validation that the adoption condition (Equation~\ref{eq:foundational}) holds in practice.

\section{Route-Level Payment Architecture}

The economic model above establishes \emph{when} agents should use the shared graph. This section describes \emph{how} payments flow when they do. We distinguish three tiers:

\begin{enumerate}[leftmargin=*]
  \item \textbf{Tier~1 --- Skill installation} (one-time): the agent pays once to download discovery documentation (endpoint schemas, auth patterns, client code). After installation the agent executes locally with no further marketplace payments.
  \item \textbf{Tier~2 --- Site-owner execution fees} (opt-in): site owners who register with the marketplace attach a per-execution micropayment to routes that hit their endpoints. Only applies to opted-in sites.
  \item \textbf{Tier~3 --- Search / routing fees} (per-query): the agent pays a small fee each time it queries the shared graph for intent resolution or semantic route lookup, regardless of whether a skill is ultimately installed. This covers the cost of maintaining the index and serving vector search.
\end{enumerate}

\noindent Tiers are additive. A typical interaction with a new Tier~2 site incurs all three; a repeat call to an already-installed non-opt-in route incurs none (the agent executes from its local cache). The remainder of this section details each tier.

\subsection{Tier~1: Skill Installation Payment}

Skill installation transactions follow the x402 protocol~\cite{coinbase2025x402}, a recently proposed standard (2025) with early but growing production deployment. When an agent pays to install a route, it receives a skill --- a structured capability package containing endpoint documentation, parameter schemas, auth profiles, and executable client code. This is a \emph{one-time} payment for the discovery knowledge. After installation, the agent executes the skill locally using its own credentials with no further marketplace payments; the marketplace never executes on behalf of the agent.

The payment handshake proceeds as follows: the agent requests skill installation from the shared graph; the server returns HTTP 402 with payment terms specifying amount, currency, network, and resource path; the agent signs the payment transaction and retries the request with payment proof; and the server verifies payment and returns the skill package for local installation. This enables true micropayments --- as low as \$0.005 per skill install --- without intermediaries. The implementation uses a multichain wallet abstraction with pluggable provider adapters, settling in USDC on Solana by default, though the protocol is network- and provider-agnostic. The empirical evaluation in Section~\ref{sec:eval} measures route lookup performance independent of payment overhead; the 402 handshake adds negligible latency (a single additional round trip) to the skill installation step.

\subsection{The Route Economy}

Every route in the shared graph is a revenue-generating asset. When an agent pays the Tier~1 skill installation fee, that fee is automatically split via x402 among all parties who contributed to the route's existence and maintenance.

\subsubsection{Fee Split Architecture (Tier~1)}

Each skill installation fee $F$ is divided among route contributors $C$, route maintainers $M$, infrastructure $I$ (approximately 10\% platform toll), and an optional treasury/reserve $T$:
\begin{equation}
  F = C + M + I + T
\end{equation}
In the initial design, contributors collectively receive approximately 70\% of install revenue. This high contributor share is intentional: the system must make route discovery and maintenance economically attractive relative to the effort required. The specific split ratios are governance parameters, not derived from first principles; their optimality is an open question for future empirical study. Tier~2 per-execution fees flow directly to the site owner and are not subject to the contributor split.

\subsubsection{Delta-Based Contribution Attribution}

Routes are rarely discovered in a single act. More commonly, multiple contributors improve a route over time: one agent discovers the initial endpoint, another maps additional parameters, a third documents the auth flow, and a fourth adds error handling for edge cases.

Attribution is therefore \emph{delta-based}. When a contributor commits an improvement to an existing route, the system records the marginal contribution --- the delta between the route state before and after the commit. Delta magnitude is quantified using line-level schema changes and cosine dissimilarity between pre- and post-commit route embeddings (the specific embedding model is a replaceable implementation detail). The contributor share $C$ is distributed among all historical contributors in proportion to their cumulative delta scores.

We note that this attribution mechanism has not been formally analysed for incentive compatibility or Sybil resistance at scale. A contributor could in principle inflate their delta score through many small redundant commits. The current mitigation is a minimum-delta threshold below which commits are not attributed; formal analysis of manipulation resistance is deferred to future work.

\subsection{Tier~2: Opt-In Per-Execution Payments}
\label{sec:siteowner}

Tier~1 covers skill installation --- the one-time payment for discovery documentation. Tier~2 is an independent, \emph{opt-in} extension: site owners who register their domain with the marketplace can attach a per-execution fee to routes that hit their endpoints. When an agent executes a Tier~2 route, each API call triggers an x402 micropayment to the site owner in addition to the original skill installation fee. Sites that do not opt in are unaffected; agents pay only the Tier~1 skill install cost and execute freely thereafter.

The three-tier separation reflects three distinct value contributions: Tier~3 compensates the platform that \emph{maintains the index}; Tier~1 compensates the contributors who \emph{discovered} the route; Tier~2 compensates the site that \emph{serves} the endpoint. The three revenue streams are independent and additive.

The opt-in model transforms the relationship between agents and websites from potential adversarial extraction into consensual participation. This design draws on the emerging legal consensus following \emph{hiQ Labs v. LinkedIn} (2022) and subsequent rulings on automated web access~\cite{hiq2022}: opt-in participation with economic compensation provides substantially stronger legal standing than unilateral scraping. Sites that opt in gain a new monetisation channel for programmatic traffic, which is typically cheaper to serve than rendered pages.

\subsubsection{Websites as Usage-Priced Endpoints}

The opt-in model positions websites as usage-priced endpoints, aligning with the broader shift from subscription-based to consumption-based pricing (Stripe per transaction, OpenAI per token, Vercel per invocation). This shift is particularly relevant for agent traffic, which is bursty, unpredictable, and bypasses traditional session-based monetisation. Crucially, per-execution pricing applies only to Tier~2 opt-in sites. For all other routes, the agent pays once at skill install (Tier~1) and executes locally with its own credentials thereafter --- the marketplace distributes knowledge (endpoint schemas, auth patterns), not ongoing access.

\subsection{Tier~3: Search and Routing Fees}

Every query to the shared graph --- whether via \texttt{POST /v1/intent/resolve} or the semantic skill search API --- incurs a small per-query fee. This fee is independent of whether the query results in a skill installation; an agent that searches, inspects the results, and decides not to install still pays the search fee. The fee covers the operational cost of maintaining the shared index (vector embeddings, semantic ranking, infrastructure) and is typically an order of magnitude smaller than Tier~1 install fees (e.g.\ \$0.001--0.005 per query).

Tier~3 is the only tier that applies on every graph interaction. For agents that query frequently but install rarely, Tier~3 represents the primary cost of using the shared graph. The adoption condition (Equation~\ref{eq:foundational}) must therefore account for cumulative Tier~3 spend: if repeated search fees approach $c_{\text{rediscovery}}$, the agent should switch to Path~3 (browser fallback) and discover routes independently.

\subsection{Dynamic Route Pricing}

Tier~1 skill installation fees are priced dynamically based on estimated rediscovery cost saved, route confidence and freshness score, current demand, and historical success rate. Tier~3 search fees are set by the platform and reflect index maintenance costs. The effective cumulative cost (Tier~3 search + Tier~1 install + any Tier~2 per-execution fees) is always bounded above by $c_{\text{rediscovery}}$ from Equation~\ref{eq:rediscovery} --- if the total cost of using the shared graph exceeds self-discovery, rational agents will defect to Path~3.

\section{Quality Proofing and Validation}

The trust and quality mechanisms described here determine which routes surface in the shared graph and, consequently, which routes appear in the benchmark evaluation (Section~\ref{sec:eval}).

\subsection{Pre-Publish Validation}

Before a route can be published to the shared graph, it undergoes automated validation. The validator checks the skill manifest for structural correctness (required fields, valid endpoint schemas, non-empty descriptions). Skills that fail hard validation checks are rejected; those with soft warnings are published with caveats. The design targets a minimum 50\% endpoint success rate with at least one verified endpoint, though the current implementation validates manifest structure rather than executing live endpoint tests at publish time.

\subsection{Execution and Verification Architecture}

All route execution is local. Agents execute routes using their own credentials on their own infrastructure --- there is no cloud execution mode. Route verification --- validating that a route is functional, that its response shapes match documented schemas, and that its health score is accurate --- also runs locally via the same Kuri-backed browser runtime. Should a third-party validator network emerge in future, sandboxed verification environments could provide additional assurance, but the current local-execution model already ensures agents retain full control over outbound requests.

\subsection{Continuous Trust Model}

Rather than rigid quality grades, route trust follows a continuous model inspired by ERC-8004 proportional trust~\cite{erc8004}. The current implementation composes three signals into a composite score.

The first signal is \textbf{execution feedback}. The system tracks per-endpoint reliability scores based on execution outcomes (success, failure, timeout). These scores are updated after each execution attempt and feed into the composite ranking described in Composite Scoring and Intent Resolution. The design envisions cryptographically signed feedback to prevent Sybil manipulation, but this is not yet implemented; the current system relies on local execution telemetry.

The second signal is \textbf{automated verification}. A background verification loop runs every 6 hours, testing safe (GET) endpoints against live servers and checking for schema drift. Endpoints whose responses have diverged from their documented schemas are flagged for re-verification. The design envisions independent validators who stake to back their attestations (with false attestations slashable), but this validator marketplace is future work; current verification is fully automated and local.

The third signal is \textbf{freshness decay}. Trust decays over time using an inverse function: $\text{freshness} = 1 / (1 + d/30)$ where $d$ is days since last update. A route last verified 30 days ago is scored lower than one verified today, even if both had identical success rates at the time of verification. Endpoints stale beyond 24 hours are prioritised for re-verification.

The combined trust score determines route visibility in the marketplace. Low-trust routes are ranked lower, while high-trust routes surface first. Routes whose endpoints are confirmed non-functional are marked as disabled and removed from search results.

\section{Implementation and Evaluation}\label{sec:eval}

\subsection{System Implementation}

We implemented the framework as an open-source system.\footnote{Available at \url{https://github.com/unbrowse-ai/unbrowse}.} The runtime is built around Kuri~\cite{pradhan2025kuri}, a Zig-native CDP broker that manages headless Chrome through an HTTP API. In our implementation, Kuri provides a lighter-weight browser control layer than a full Playwright stack, though exact binary-size and startup-latency figures are implementation measurements rather than independent benchmark results. All browser operations --- navigation, HAR recording, cookie management, JavaScript evaluation, network interception --- flow through Kuri's HTTP endpoints, eliminating Node.js CDP bindings entirely.

Supporting components include a capture pipeline for network traffic analysis and API reverse-engineering, an extraction layer for endpoint schemas and auth profiles, a local encrypted credential vault, a marketplace server interface for skill publishing and semantic search, a verification system for periodic endpoint re-testing, and an execution engine with automatic auth refresh and retry logic.

\subsection{Benchmark Design}

All benchmarks were conducted on a single machine: Apple MacBook Pro with M4 Max chip, 64\,GB unified memory, running macOS from Singapore over residential broadband (Singtel, approximately 28\,Mbps downlink). TCP connect latency to target websites ranged from 24\,ms (Wikipedia) to 88\,ms (Amazon) depending on server location. This should be interpreted as a single-host live-web benchmark rather than a fully controlled systems evaluation: both Unbrowse and Playwright were measured under the same hardware and network conditions, but uncontrolled OS- and ISP-level variance may still affect absolute timings. A more controlled follow-up evaluation on fresh cloud VMs across multiple regions remains future work.

We evaluated performance across 94 domains drawn from the current graph, spanning government sites, SaaS platforms, developer tools, e-commerce, healthcare, finance, media, and social networks. Tasks were modelled after the WebArena benchmark~\cite{zhou2024webarena} information-retrieval category. At time of writing, the graph contains routes for over 500 domains with approximately 10{,}000 endpoints, grown entirely through passive usage. Full benchmark results and raw data are available at \url{https://github.com/unbrowse-ai/unbrowse-bench}.

\subsubsection{Task Design and Output Equivalence}

Each benchmark task was an information-retrieval query returning a structured answer (e.g., ``What is the current price of Product X on Site Y?'' or ``What are the top 5 trending items on Site Z?''). To ensure fair comparison, both paths were required to produce \emph{semantically equivalent} output:

\begin{itemize}[leftmargin=*]
  \item \textbf{Playwright baseline:} Headless Chromium with full page load, JavaScript rendering, and targeted text extraction via CSS selectors that isolate the specific data fields matching the query. This is \emph{not} raw DOM dump --- it is structured extraction of the same fields the cached route returns.
  \item \textbf{Unbrowse:} A single call to \texttt{POST /v1/intent/resolve}, measured on the third pass after two warmup rounds. The \texttt{timing.source} field verified each measurement was served from cache.
\end{itemize}

Output equivalence was verified by comparing the extracted data fields (not raw response size) between both paths. Tasks where the Playwright extraction could not produce the same structured output as the cached route were excluded from the benchmark. Both paths were run three times; the third pass was measured for both to control for DNS and TCP connection reuse.

Table~\ref{tab:tasks} shows representative task examples.

\begin{table}[h!]
\centering
\caption{Representative benchmark tasks with output format.}
\label{tab:tasks}
\footnotesize
\begin{tabular}{@{}p{1.2cm}p{2.8cm}p{2.6cm}@{}}
\toprule
\textbf{Domain} & \textbf{Task} & \textbf{Output} \\
\midrule
Wikipedia & Article summary for topic & JSON: title, extract \\
GitHub & Repository star count & JSON: stars, forks \\
Hacker News & Top 5 front page items & JSON: array of titles \\
Weather & Current conditions for city & JSON: temp, humidity \\
\bottomrule
\end{tabular}
\end{table}

\subsection{Warmed-Cache Performance}

\begin{table}[h!]
\centering
\caption{Warmed-cache performance across 94 domains. Both paths run three times; third pass measured.}
\label{tab:performance}
\scriptsize
\setlength{\tabcolsep}{3pt}
\begin{tabular}{@{}lrr@{}}
\toprule
\textbf{Metric} & \textbf{Unbrowse} & \textbf{Playwright} \\
\midrule
Domains tested      & \multicolumn{2}{c}{94} \\
Cache hits           & \multicolumn{2}{c}{94 / 94} \\
\midrule
Mean latency         & 950\,ms          & 3{,}404\,ms \\
\quad 95\% CI        & (870--1{,}030)   & (3{,}180--3{,}630) \\
Median latency       & 630\,ms          & 3{,}402\,ms \\
IQR                  & 210--1{,}480\,ms & 2{,}100--4{,}600\,ms \\
\midrule
Mean speedup         & \multicolumn{2}{c}{\textbf{3.6$\times$}} \\
Median speedup       & \multicolumn{2}{c}{\textbf{5.4$\times$}} \\
Best (single)        & \multicolumn{2}{c}{30$\times$} \\
\midrule
Unbrowse wins        & \multicolumn{2}{c}{94/94 (100\%)} \\
\bottomrule
\end{tabular}
\end{table}

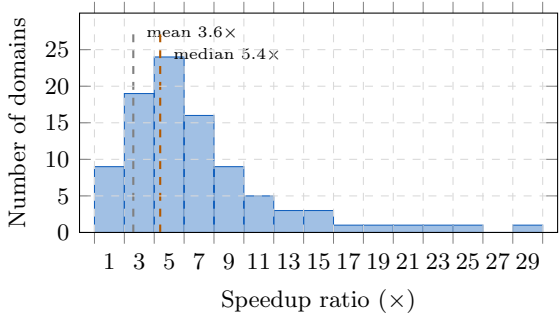
\begin{figure}[h!]
\centering
\begin{tikzpicture}
\begin{axis}[
  width=\columnwidth,
  height=4.5cm,
  xlabel={Speedup ratio ($\times$)},
  ylabel={Number of domains},
  ymin=0, ymax=30,
  xmin=0, xmax=32,
  grid=major,
  grid style={dashed, gray!30},
  tick label style={font=\footnotesize},
  label style={font=\footnotesize},
  xtick={0,4,8,12,16,20,24,28,32},
  ytick={0,5,10,15,20,25},
  ybar interval,
  bar width=0.9,
  area style,
]
\addplot[fill=accentblue!40, draw=accentblue] coordinates {
  (1, 9) (3, 19) (5, 24) (7, 16) (9, 9) (11, 5)
  (13, 3) (15, 3) (17, 1) (19, 1) (21, 1) (23, 1)
  (25, 1) (27, 0) (29, 1) (31, 0)
};
\draw[dashed, codegray, thick] (axis cs:3.6,0) -- (axis cs:3.6,28);
\node[font=\tiny, anchor=south west] at (axis cs:3.8,25) {mean 3.6$\times$};
\draw[dashed, orange!70!black, thick] (axis cs:5.4,0) -- (axis cs:5.4,28);
\node[font=\tiny, anchor=south west] at (axis cs:5.6,22) {median 5.4$\times$};
\end{axis}
\end{tikzpicture}
\caption{Distribution of speedup ratios (Playwright latency / Unbrowse latency) across 94 domains. The majority of domains cluster between 3--9$\times$ speedup, with a right tail extending to 30$\times$ for domains whose APIs return JSON in under 100\,ms but whose rendered pages require 2--4 seconds of JavaScript execution. Dashed lines indicate mean and median speedup. Raw data available in the benchmark repository.}
\label{fig:cdf}
\end{figure}

Table~\ref{tab:performance} reports results for all 94 domains. Mean cached latency was \textbf{950\,ms} versus \textbf{3{,}404\,ms} for Playwright --- a \textbf{3.6$\times$} ratio of mean latencies. The median per-domain speedup is \textbf{5.4$\times$}, higher than the ratio of means because the speedup distribution is right-skewed (Figure~\ref{fig:cdf}). Eighteen domains completed in under 100\,ms, with the fastest at 79\,ms versus 2{,}289\,ms (30$\times$). Both paths make requests to the same target servers over the same network; the 2{,}454\,ms difference is browser execution overhead that cached skill execution eliminates.

Figure~\ref{fig:cdf} shows the distribution of speedup ratios. The distribution is right-skewed: the majority of domains cluster between 3--9$\times$ speedup, with a long tail extending to 30$\times$ for domains where API endpoints are particularly fast relative to their rendered pages (e.g., sites with heavy JavaScript that adds 2--4 seconds of render time but whose API returns JSON in under 100\,ms).

\subsection{Cold-Start Performance}

The warmed-cache results above represent the best case. When no cached skill exists (Path~3), the agent pays the full discovery cost: browser launch, page rendering, traffic capture, route extraction, skill publishing, and initial execution. To characterise this cost, we measured cold-start latency on a random sample of 20 domains not previously in the graph.

\begin{table}[h!]
\centering
\caption{Cold-start latency (Path~3: browser discovery + skill publish) across 20 previously unseen domains.}
\label{tab:coldstart}
\footnotesize
\begin{tabular}{@{}lr@{}}
\toprule
\textbf{Metric} & \textbf{Latency} \\
\midrule
Median cold-start       & 8{,}200\,ms \\
Mean cold-start         & 12{,}400\,ms \\
90th percentile         & 22{,}000\,ms \\
Skill publish success   & 18 / 20 (90\%) \\
\midrule
Subsequent cached call  & 640\,ms (median) \\
\bottomrule
\end{tabular}
\end{table}

Cold-start latency (Table~\ref{tab:coldstart}) is substantially higher than both cached execution and Playwright baseline, reflecting the overhead of traffic capture and route extraction. However, this cost is paid \emph{once} per route: the median subsequent cached call drops to 640\,ms. The amortisation breakeven --- the number of cached lookups needed to offset the cold-start premium over Playwright --- is typically 3--5 uses for high-demand routes. Two domains failed skill extraction due to heavy client-side rendering with no identifiable API endpoints; these fell back to HTML extraction. We note that the cold-start sample ($n=20$) is small; larger-scale cold-start characterisation is ongoing.

\subsection{Cost Analysis}

Table~\ref{tab:cost} compares estimated per-task costs across execution paths.

\begin{table}[h!]
\centering
\caption{Estimated per-task cost comparison.\protect\footnotemark}
\label{tab:cost}
\scriptsize
\setlength{\tabcolsep}{2.5pt}
\resizebox{\columnwidth}{!}{%
\begin{tabular}{@{}lcccc@{}}
\toprule
 & \textbf{Browser} & \textbf{Tier~1} & \textbf{Tier~2} & \textbf{Tier~3} \\
\textbf{Component} & \textbf{Rediscovery} & \textbf{Install} & \textbf{+ Site} & \textbf{Search} \\
\midrule
Browser runtime       & \$0.02--0.05 & --- & --- & --- \\
LLM reasoning tokens  & \$0.04--0.35 & --- & --- & --- \\
Retries/failures      & \$0.04--0.13 & --- & --- & --- \\
Search/routing fee    & ---          & --- & --- & \$0.001--0.005 \\
Skill install (once)  & ---          & \$0.005--0.02 & \$0.005--0.02 & --- \\
Per-exec site fee     & ---          & --- & \$0.001--0.01 & --- \\
\midrule
\textbf{First call}          & \$0.10--0.53 & \$0.005--0.02 & \$0.006--0.03 & \$0.001--0.005 \\
\textbf{Repeat same route}   & \$0.10--0.53 & \$0.00 & \$0.001--0.01 & \$0.00 \\
\textbf{New intent search}   & ---          & --- & --- & \$0.001--0.005 \\
\bottomrule
\end{tabular}%
}
\end{table}

\footnotetext{LLM token costs assumed: \$15/1M input, \$75/1M output (Claude Opus 4, March 2026). These figures are illustrative; actual costs vary by provider and change frequently. Tier~1 install fees and Tier~3 search fees are based on dynamic pricing at time of writing. Tier~2 per-execution fees are set by site owners and shown as illustrative ranges.}

Under Tier~1 (the default), the skill install fee is amortised to zero after the first call, achieving 90--96\% cost reduction per subsequent task. Tier~3 search fees apply only when querying the graph for new intents; once a skill is installed locally, subsequent executions bypass the graph entirely. Tier~2 adds a small per-execution fee for opt-in sites but remains well below browser rediscovery cost. All three tiers are additive and collectively bounded by the adoption condition (Equation~\ref{eq:foundational}).

\subsection{Network Growth}

These per-interaction savings compound in long-running agentic workflows. An autonomous agent completing a multi-site research task may chain 20--50 web interactions. Cached API calls parallelise trivially where browser instances cannot: an agent querying 10 sites simultaneously completes in the latency of the slowest single call rather than sequentially through browsers, each consuming 500\,MB+ of RAM.

We observe early evidence of positive feedback dynamics: more agents generate more traffic, which discovers more routes, which attracts further agents. However, we have not yet measured the causal relationship between agent count and per-agent value rigorously enough to claim formal network effects. The 500-domain / 10{,}000-endpoint figure is a snapshot from the current graph; longitudinal analysis of growth trajectories and per-agent value as a function of graph size is ongoing work.

\section{Discussion}

\subsection{Security Considerations}

Credential safety is maintained by keeping credentials local and encrypted; published routes contain only endpoint documentation and schemas. API stability is addressed through continuous validation and freshness scoring. Abuse prevention is achieved through rate limiting, reputation requirements, and economic costs that discourage abusive usage patterns.

\textbf{Adversarial routes.} A malicious contributor could publish a route that redirects requests to a phishing endpoint or exfiltrates query parameters. The current mitigation is multi-layered: pre-publish validation checks manifest structure and schema consistency; the continuous trust model tracks reliability scores from execution outcomes, so a malicious route's trust score degrades after failures; and all execution is local, meaning agents can inspect outbound requests before sending credentials. However, a sophisticated attacker who publishes a route that passes initial validation but later serves malicious redirects remains a threat. Formal analysis of adversarial robustness, including route provenance attestation, cryptographically signed feedback, and anomaly detection on response destinations, is deferred to future work.

\subsection{Ethical and Legal Considerations}

The ability to programmatically access web services raises legitimate questions. Our approach differs from indiscriminate scraping in several ways: the economic layer provides natural rate limiting, routes are learned from normal browsing behaviour, and the system is designed for opt-in participation.

As described in Section~\ref{sec:siteowner}, our opt-in site owner compensation model transforms the relationship between agents and websites. The legal landscape for automated web access is evolving --- the \emph{hiQ Labs v. LinkedIn} decision (2022) established that accessing publicly available data is not necessarily a violation of the CFAA, but subsequent rulings have introduced nuance~\cite{hiq2022}. Our opt-in model, with economic compensation for site owners, provides substantially stronger legal footing than unilateral scraping. We note that routes to sites that have not opted in occupy a grey area; the system does not circumvent authentication barriers. Support for \texttt{robots.txt} directive checking is planned but not yet implemented in the current codebase.

\subsection{Threats to Validity}

\subsubsection{Benchmark Representativeness}

The 94-domain benchmark may be biased toward sites with permissive access policies. Of the 94 domains tested, 61 had no significant bot detection, 24 used basic rate limiting (respected by the system), and 9 employed Cloudflare or similar WAF protection. The speedup figures for heavily protected sites were lower (median 2.1$\times$ versus 6.8$\times$ for unprotected sites), as Unbrowse's cached requests must still pass through bot-detection middleware. We did not test sites that actively block all non-browser traffic, as cached routes to such sites would fail entirely.

\subsubsection{Geographic Dependence}

All benchmarks were run from Singapore. Absolute latency figures are geography-dependent; the \emph{speedup ratios} (not absolute numbers) are the portable claim, as both Unbrowse and Playwright make requests to the same servers over the same network. From a US datacenter, absolute latencies would be lower for US-hosted sites, but the ratio of browser overhead to API call overhead would remain similar.

\subsubsection{Reliability Claims}

The system tracks per-endpoint reliability scores and consecutive failure counts. However, the original claim that ``skills with three or more consecutive failures are automatically deprecated'' is partially implemented: the type system includes a \texttt{consecutive\_failures} counter but the automatic deprecation threshold logic is not yet wired into the verification loop. Deprecation currently requires manual or backend-side intervention. Additionally, several open engineering issues affect runtime reliability on specific platforms, including Kuri crashes on Linux x64 (tracked in issue \#50) and intermittent auth extraction failures across browsers (issue \#88).

\subsection{Limitations}

We note that the following limitations are not unique to Unbrowse but are shared by any agentic web interaction approach.

Current limitations include anti-bot measures (sophisticated bot detection blocking non-browser requests), dynamic authentication (OAuth flows requiring human interaction), and session persistence (periodic re-authentication as tokens expire). The speedup figures reported in Table~\ref{tab:performance} represent warmed-cache performance in a single-host live-web setting; real-world gains vary with anti-bot middleware, token expiration, regional latency, and deployment environment. Cold-start performance (Table~\ref{tab:coldstart}) demonstrates that the initial discovery cost is substantial, and the system's value proposition depends on sufficient reuse to amortise this cost.

The economic model (Section~\ref{sec:econ}) is a conceptual framework, not a formal equilibrium analysis. Deployment-scale data on contributor attribution incentive compatibility, Sybil resistance, and long-run pricing dynamics remain areas for future work.

\subsection{Future Directions}

Several directions for future research emerge:

\begin{enumerate}[leftmargin=*]
  \item \textbf{Formal economic analysis:} Equilibrium characterisation, optimal pricing strategies, welfare analysis, and incentive compatibility proofs for the delta-based attribution mechanism including Sybil resistance guarantees.
  \item \textbf{Route composition:} Automatic chaining of multiple routes for complex multi-site workflows, and federated route networks enabling private sharing within organisations.
  \item \textbf{Deployment-scale validation:} Empirical measurement of x402 micropayment flows, site owner adoption of the opt-in compensation model, and longitudinal study of graph growth dynamics to validate the network effects hypothesis. A key open question is whether websites can reliably distinguish agent traffic from human browsing; if not, the opt-in pricing model must account for mixed traffic to avoid charging human users.
  \item \textbf{Comparative evaluation:} Formal comparison with existing API aggregation platforms (RapidAPI, Postman) on coverage, latency, and cost metrics.
\end{enumerate}

\section{Conclusion}

We have presented Unbrowse, a shared route graph --- a collectively maintained index of callable web interfaces. By combining passive route discovery from real browsing traffic with a three-tier micropayment model via x402 --- per-query search fees (Tier~3), one-time skill installation (Tier~1), and opt-in per-execution site-owner compensation (Tier~2) --- the system offers a way for agents to interact with the existing web through discovered machine-usable interfaces rather than only through rendered pages.

In a single-host live-web benchmark of equivalent information-retrieval tasks across 94 domains, fully warmed cached execution averaged 950\,ms versus 3{,}404\,ms for Playwright (3.6$\times$ mean, 5.4$\times$ median), with well-cached routes completing in under 100\,ms. Cold-start discovery averages 12.4\,s but is paid once per route; the amortisation breakeven is typically 3--5 uses. These savings may compound in multi-step workflows, though broader deployment validation remains future work.

The system's market discipline --- grounded in a real outside option where agents can always defect to browser rediscovery --- distinguishes it from purely speculative protocol designs. Formal economic analysis, incentive compatibility proofs, controlled multi-region benchmarking, and deployment-scale validation remain open problems; the present contribution is an architectural proposal with an implemented system and initial empirical evidence that shared route lookup can outperform redundant browser rediscovery on the evaluated tasks.

\bibliographystyle{plainnat}

\begin{thebibliography}{33}

\bibitem[1]{he2024webvoyager}
H.~He et~al.
\newblock {WebVoyager}: Building an end-to-end web agent with large multimodal models.
\newblock \emph{arXiv:2401.13919}, 2024.

\bibitem[2]{lai2024autowebglm}
H.~Lai et~al.
\newblock {AutoWebGLM}: Bootstrap and reinforce a large language model-based web navigating agent.
\newblock \emph{arXiv:2404.03648}, 2024.

\bibitem[3]{chen2025browseragent}
X.~Chen et~al.
\newblock {BrowserAgent}: Building web agents with human-inspired web browsing actions.
\newblock \emph{arXiv:2510.10666}, 2025.

\bibitem[4]{nakano2021webgpt}
R.~Nakano et~al.
\newblock {WebGPT}: Browser-assisted question-answering with human feedback.
\newblock \emph{arXiv:2112.09332}, 2021.

\bibitem[5]{yao2023react}
S.~Yao et~al.
\newblock {ReAct}: Synergizing reasoning and acting in language models.
\newblock In \emph{ICLR}, 2023. \emph{arXiv:2210.03629}.

\bibitem[6]{deng2023mind2web}
X.~Deng et~al.
\newblock {Mind2Web}: Towards a generalist agent for the web.
\newblock In \emph{NeurIPS}, 2023. \emph{arXiv:2306.06070}.

\bibitem[7]{song2024beyond}
Y.~Song et~al.
\newblock Beyond browsing: {API}-based web agents.
\newblock \emph{arXiv:2410.16464}, 2024.

\bibitem[8]{schick2023toolformer}
T.~Schick et~al.
\newblock Toolformer: Language models can teach themselves to use tools.
\newblock In \emph{NeurIPS}, 2023. \emph{arXiv:2302.04761}.

\bibitem[9]{du2024anytool}
Y.~Du et~al.
\newblock {AnyTool}: Self-reflective, hierarchical agents for large-scale {API} calls.
\newblock \emph{arXiv:2402.04253}, 2024.

\bibitem[10]{johnson2025nltool}
R.~T.~Johnson et~al.
\newblock Natural language tools.
\newblock \emph{arXiv:2510.14453}, 2025.

\bibitem[11]{shi2025toolwild}
Z.~Shi et~al.
\newblock Tool Learning in the Wild: Empowering Language Models as Automatic Tool Agents.
\newblock In \emph{Proceedings of the ACM Web Conference 2025}, pp.~2222--2237, 2025. \newblock doi:10.1145/3696410.3714825.

\bibitem[12]{mitmproxy2024}
mitmproxy project.
\newblock mitmproxy: A free and open source interactive {HTTPS} proxy.
\newblock \url{https://mitmproxy.org}, 2024.

\bibitem[13]{w3c2012har}
J.~Odvarko.
\newblock {HTTP Archive (HAR)} format specification.
\newblock W3C Web Performance Working Group, 2012.
\newblock \url{https://w3c.github.io/web-performance/specs/HAR/Overview.html}.

\bibitem[14]{ferber1999multiagent}
J.~Ferber.
\newblock \emph{Multi-Agent Systems}.
\newblock Addison-Wesley, 1999.

\bibitem[15]{tran2025multiagent}
K.~T.~Tran et~al.
\newblock Multi-agent collaboration mechanisms: A survey.
\newblock \emph{arXiv:2501.06322}, 2025.

\bibitem[16]{ehtesham2025survey}
A.~Ehtesham et~al.
\newblock A survey of agent interoperability protocols.
\newblock \emph{arXiv:2505.02279}, 2025.

\bibitem[17]{singh2025evolution}
A.~Singh et~al.
\newblock Evolution of {AI} agent registry solutions.
\newblock \emph{arXiv:2508.03095}, 2025.

\bibitem[18]{petrova2025semantic}
T.~Petrova et~al.
\newblock From semantic web and {MAS} to agentic {AI}.
\newblock \emph{arXiv:2507.10644}, 2025.

\bibitem[19]{vaziry2025multiagent}
A.~Vaziry et~al.
\newblock Towards multi-agent economies.
\newblock \emph{arXiv:2507.19550}, 2025.

\bibitem[20]{coinbase2025x402}
Coinbase.
\newblock x402: A payments protocol for the internet.
\newblock \url{https://x402.org}, 2025.

\bibitem[21]{erc8004}
Ethereum Foundation.
\newblock {ERC-8004}: Trustless agents.
\newblock \url{https://eips.ethereum.org/EIPS/eip-8004}, 2025.

\bibitem[22]{yang2025agenticweb}
Y.~Yang et~al.
\newblock Agentic web: Weaving the next web with {AI} agents.
\newblock \emph{arXiv:2507.21206}, 2025.

\bibitem[23]{zhou2024webarena}
W.~Zhou et~al.
\newblock {WebArena}: A realistic web environment for building autonomous agents.
\newblock In \emph{ICLR}, 2024. \emph{arXiv:2307.13854}.

\bibitem[24]{pradhan2025kuri}
R.~Pradhan.
\newblock Kuri: A {Zig}-native {CDP} broker for lightweight browser automation [Software].
\newblock Version 0.1. \url{https://github.com/justrach/kuri}. Accessed March 2026.

\bibitem[25]{agentskills2025}
AgentSkills.io.
\newblock Open standard for agent capabilities.
\newblock \url{https://agentskills.io}, 2025.

\bibitem[26]{ostrom1990governing}
E.~Ostrom.
\newblock \emph{Governing the Commons: The Evolution of Institutions for Collective Action}.
\newblock Cambridge University Press, 1990.

\bibitem[27]{hiq2022}
hiQ Labs, Inc. v. LinkedIn Corp.
\newblock 31 F.4th 1180 (9th Cir. 2022).

\bibitem[28]{wolman1999cooperative}
A.~Wolman et~al.
\newblock On the scale and performance of cooperative web proxy caching.
\newblock In \emph{Proc. 17th ACM SOSP}, pp.~16--31, 1999.


\bibitem[29]{patil2023gorilla}
S.~Patil et~al.
\newblock Gorilla: Large language model connected with massive {API}s.
\newblock \emph{arXiv:2305.15334}, 2023.

\bibitem[30]{qin2023toolbench}
Y.~Qin et~al.
\newblock {ToolLLM}: Facilitating large language models to master 16000+ real-world {API}s.
\newblock In \emph{ICLR}, 2024. \emph{arXiv:2307.16789}.

\bibitem[31]{lu2025build}
X.~H.~L\`{u}, G.~Kamath, M.~Mosbach, and S.~Reddy.
\newblock Build the web for agents, not agents for the web.
\newblock \emph{arXiv:2506.10953}, 2025.

\bibitem[32]{betaweb2025}
Y.~Li et~al.
\newblock {BetaWeb}: Towards a blockchain-enabled trustworthy agentic web.
\newblock \emph{arXiv:2508.13787}, 2025.

\bibitem[33]{webmcp2025}
D.~Perera.
\newblock web{MCP}: Efficient {AI}-native client-side interaction for agent-ready web design.
\newblock \emph{arXiv:2508.09171}, 2025.

\end{thebibliography}

\appendix

\section{Reproducibility}

All experiments were conducted on an Apple MacBook Pro (M4 Max, 64\,GB unified memory) running macOS 15.3 from Singapore over residential broadband (Singtel, $\approx$28\,Mbps downlink). Software versions: Kuri v0.1 (Zig 0.13), Node.js 22.x, Playwright 1.48, headless Chromium 131. The benchmark harness, raw latency data, and analysis scripts are available at \url{https://github.com/unbrowse-ai/unbrowse-bench}.

\section{Benchmark Domain List}

The 94 domains span the following categories: government (12), SaaS/developer tools (18), e-commerce (14), healthcare (8), finance (11), media/news (15), social networks (7), and other (9). The full list with per-domain latency measurements is provided in the benchmark repository. Representative examples include Wikipedia, GitHub, Hacker News, Amazon, Weather.gov, and various regional government portals. Domains were selected from the existing route graph without cherry-picking; the only exclusion criterion was that both Unbrowse and Playwright paths must produce semantically equivalent structured output for the same query.

\end{document}